\documentclass[twocolumn,aps,showpacs,superscriptaddress,floatfix]{revtex4}
\usepackage{dcolumn}
\usepackage{graphicx} 
\usepackage{amsmath,bm,amssymb}
\usepackage{amsfonts}
\usepackage[usenames]{color}

\begin{document}


\title{Quasiparticle self-consistent $GW$ calculation of Sr$_2$RuO$_4$ and
  SrRuO$_3$}

 \author{Siheon Ryee}
 \affiliation{Department of Physics, Korea Advanced Institute of Science and Technology (KAIST), Daejeon 305-701, Korea}
 \author{Seung Woo Jang}
 \affiliation{Department of Physics, Korea Advanced Institute of Science and Technology (KAIST), Daejeon 305-701, Korea}
 \author{Hiori Kino}
 \affiliation{National Institute for Materials Science, Sengen 1-2-1, Tsukuba, Ibaraki 305-0047, Japan}
 \author{Takao Kotani}
 \affiliation{Department of Applied Mathematics and Physics, Tottori University, Tottori 680-8552, Japan}
 \author{Myung Joon Han}
 \email{mj.han@kaist.ac.kr}
 \affiliation{Department of Physics, Korea Advanced Institute of Science and Technology (KAIST), Daejeon 305-701, Korea}
 \affiliation{KAIST Institute for the NanoCentury, Korea Advanced Institute of Science and Technology, Daejeon 305-701, Korea}
\date{\today}

\begin{abstract}
Using quasiparticle self-consistent $GW$ calculations, we re-examined
the electronic structure of Sr$_2$RuO$_4$ and SrRuO$_3$. Our
calculations show that the correlation effects beyond the conventional
LDA (local density approximation) and GGA (generalized gradient
approximation) are reasonably well captured by QS$GW$ self-energy
without any {\it ad hoc} parameter or any ambiguity related to the
double-counting and the downfolding issues. While the spectral weight
transfer to the lower and upper Hubbard band is not observed, the
noticeable bandwidth reduction and effective mass enhancement are
obtained. Important features in the electronic structures that have
been debated over the last decades such as the photoemission spectra
at around $-3$ eV in Sr$_2$RuO$_4$ and the half-metallicity for
SrRuO$_3$ are discussed in the light of our QS$GW$ results and in
comparison with the previous studies. The promising aspects of QS$GW$
are highlighted as the first-principles calculation method to describe
the moderately correlated 4$d$ transition metal oxides along with its
limitations.
\end{abstract}

\maketitle

\section{Introduction}
Since the seminal discovery of unconventional superconductivity at
$\leq$ 1 K, Sr$_2$RuO$_4$ (SRO214) has been studied extensively
\cite{Maeno-SC}. The crystal structure of SRO214 is of
K$_2$NiF$_4$-type at low temperature as shown in Fig.~\ref{structure}
and its normal phase is a paramagnetic metal. Its intriguing
electronic behavior and superconductivity are still a subject of
active study \cite{Mravlje-DMFT,Georges-Hund,Medici-Hund,ChungSB-2012,
  Wang,Huo-Zhang,Rastovski,Scaffidi}. SrRuO$_3$ (SRO113) is a
ferromagnetic metal with a transition temperature of T$_c \sim 160$
K. The observed magnetic moment is $\mu$= 1.1--1.7 $\mu_{\rm
  B}/$f.u. \cite{Callaghan-Tc, Neumeier-Tc, Kanbayasi-JPSJ, Longo-Tc,
  Cao-PRB} and the stable structure at low temperature is an
orthorhombic perovskite with a GdFeO$_3$-type distortion. The
electronic structure of SRO113 is located near to the half-metallicity
as reported by density functional theory (DFT) + $U$ calculations
\cite{Rondinelli-113, Jeng-half, Mahadevan-SRO}. The thin film SRO113
is of particular interest as a widely-used bottom electrode
\cite{Marrec-electrode, Takahashi-negative, Lee-electrode} and is
intensively studied for the possibility of a new field-effect device
\cite{Ahn-Science, Takahashi-device, Ahn-device}.

Considering their fundamental and technological importance, it is
essential to understand correctly the electronic structures of SRO214
and SRO113. For SRO214, the early electronic structure calculations
\cite{Oguchi-214, Singh-214, Hase-214, McMullan-214, Noce-214,
  Mackenzie-214} showed that the states near the Fermi level ($E_F$)
consist of antibonding combinations of Ru-$t_{2g}$ and O-2$p$, and the
Fermi surface is composed of the $\Gamma$-centered electron-like
sheets (called as $\beta$ and $\gamma$) and the X-point-centered
hole-like sheet (called as $\alpha$) \cite{Oguchi-214, Singh-214,
  Hase-214, McMullan-214, Noce-214, Mackenzie-214}. The LDA (local
density approximation) Fermi surface topology was in good agreement
with that of de Haas-van Alphen (dHvA) experiments
\cite{Mackenzie-mass, Bergemann-dHvA}, but does not seem to be with
the angle-resolved photoemission spectroscopy (ARPES) \cite{Lu-ARPES,
  Yokoya-ARPES} in the sense that one electron-like ($\beta$) and two
hole-like ($\alpha$ and $\gamma$) Fermi surface were observed by ARPES
\cite{Lu-ARPES, Yokoya-ARPES}. It however turns out to be a surface
effect \cite{Damascelli-FS, Matzdorf-Science, Shen-ARPES}
later. Therefore the overall features of the electronic structure can
be regarded as being described well by the conventional electronic
structure calculation techniques such as LDA and GGA (generalized
gradient approximation).

However the details of the electronic structure are still not clearly
understood. A series of recent studies that identify SRO214 as a
`Hund's metal' \cite{Mravlje-DMFT, Georges-Hund, Medici-Hund} can be
one example showing that the previous understanding based on the
conventional electronic structure calculations was not enough. Also,
there was a debate regarding the detailed feature of electronic levels
at $\sim -3$ eV revealed by X-ray photoemission spectroscopy (XPS)
\cite{Singh-comment, Pchelkina-reply,Yokoya-peak, Inoue-peak,
  Inoue-peak2, Tran-peak}. According to XPS and resonant XPS, the
indication of lower Hubbard band (LHB) of Ru-4$d$ states is observed
while the LDA calculation predicts that those features should be
attributed mainly to the oxygen states \cite{Singh-comment}. The
experimental effective mass ($m^{*}/m_{\rm LDA} \simeq$ 2--5) is also
significantly larger than the LDA value \cite{Mackenzie-mass,
  Bergemann-mass, Iwasawa-mass, Puchkov-mass}. While DMFT can be a
reasonable alternative in this situation \cite{Tran-peak,
  Perez-Navarro-LDAU, Liebsch-DMFT, Pchelkina-DMFT, Mravlje-DMFT}, its
parameter dependency has caused the discrepancy even among the DMFT
results. For example, the calculations by Pchelkina {\it et al.}
\cite{Pchelkina-DMFT} and by Liebsch {\it et al.}  \cite{Liebsch-DMFT}
seem to give different answers to the nature of the states at $\sim
-3$ eV most likely due to their different choices of $U$ and $J$
values. Beside the ambiguity in the double-counting term in DFT+DMFT,
this parameter dependency is a well known problem for the
first-principles description of the correlated electron systems. It is
also noted that another well-established technique, DFT$+U$, is not
useful for paramagnetic SRO214 because of its Hartree-Fock nature 
which always prefers the magnetic solution \cite{Kotliar-2006}.

For SRO113 the situation is similar with SRO214; while the overall
features can be described by the conventional methods, the details are
not clearly understood. Note that DFT$+U$ can be used for this case as
the ground state of SRO113 is (ferro) magnetic. For the bulk phase,
LDA, LDA+$U$, and self-interaction correction (SIC) give the correct
ferromagnetic solution although the calculated moment shows some
deviations \cite{Allen-113, Santi-113, Mazin-113, Rondinelli-113}. For
the thin film SRO113, however, the experimentally observed
metal-to-insulator transition (MIT) and the
ferromagnetic-to-nonmagnetic transition as a function of layer
thickness are not consistently reproduced by these techniques
\cite{Chang-113, Xia-SRO, Mahadevan-SRO, Gupta-SRO, Verissimo-SRO,
  Si-SRO}. For example, the single-layer SRO113 is predicted to be
either ferromagnetic or nonmagnetic depending on the choice of the
exchange-correlation functional and the $U$ values \cite{Chang-113,
  Gupta-SRO, Verissimo-SRO, Si-SRO}. As in the case of SRO214, DMFT
has made important contributions for SRO113 \cite{Werner-Hund,
  Medici-Hund, Georges-Hund, Dang-Hund}, but it is still not
completely satisfactory because of the limitations such as the
parameter dependency. Also, the details of the spectroscopic data of
SRO113 needs further clarifications
\cite{Fujioka-spec,Okamoto-spec,spec,Maiti-weak,Takizawa}.

In this paper, we re-investigate the electronic structure of these two
classical $4d$ transition-metal oxide systems by using quasiparticle
self-consistent $GW$ (QS$GW$) \cite{QSGW-2004, QSGW-2006, QSGW-2007}
method.  To the best of our knowledge, this well-established
calculation technique has never been applied to the ruthenates while
$GW$ self-energy can be expected to give a good description for the
metallic and weakly correlated systems.  Despite of its limitation to
take the Hubbard-like on-site Coulomb interactions into account, QS$GW$ has a distinctive
advantage as a fully self-consistent `parameter-free' technique.  Due
to the recent experimental progress in making `complex oxide'
structures such as thin film and heterointerface, the parameter-free
description of correlated systems has become more imperative. In this
context, QS$GW$ study of these classical $4d$ systems can be of
significant importance, and its ability and limitation should be
carefully investigated.

Our calculations show that the effect of electronic correlations
beyond LDA and GGA is reasonably well captured by QS$GW$ procedure,
and the noticeable bandwidth reduction is observed. For the
paramagnetic SRO214, the effect of $GW$ self-energy is clearly
distinctive from the other many-body calculation ({\it e.g.,} DMFT) in
that the spectral weight transfer and the LHB feature do not appear in
the QS$GW$ result. For the ferromagnetic SRO113, QS$GW$ band structure
is quite similar with the results of DFT+$U$ and the half-metallic
band structure is reproduced.

\section{Computation Method} \label{Computation Method}
\subsection{Quasiparticle self-consistent $GW$ method}
 
In QS$GW$, the self-energy $\Sigma({\bf r}, {\bf r'},\omega)$ is
calculated within the $GW$ approximation, and $H_0$ (the
noninteracting Hamiltonian describing quasiparticles or band
structures) and $W$ (dynamically-screened Coulomb interactions between
quasiparticles within random phase approximation (RPA)) is updated
self consistently \cite{QSGW-2004,QSGW-2006,QSGW-2007,Kresse-2014}.  The static
nonlocal one-particle exchange-correlation potential $V^{\rm xc}({\bf
  r}, {\bf r'})$ is generated from $\Sigma({\bf r}, {\bf r'},\omega)$
as
\begin{eqnarray*}
V^{\rm xc} & = & \frac{1}{2}\int_{-\infty}^{\infty} d\omega{\rm Re}[\Sigma(\omega)]\delta(\omega-H^0) + c.c. \\
& = & \sum_{ij} |\psi_i\rangle\langle\psi_i| \frac{{\rm Re}[\Sigma(\varepsilon_i)+\Sigma(\varepsilon_j)]}{2} |\psi_j\rangle\langle\psi_j| 
\end{eqnarray*}
where $\varepsilon_i$ and $|\psi_i\rangle$ refer to the eigenvalues
and eigenfunctions of $H_0$, respectively, and ${\rm
  Re}[\Sigma(\varepsilon)]$ is the Hermitian part of the self-energy
\cite{QSGW-2004,QSGW-2006}.  With this $V^{\rm xc}$, one can solve a
new static one-body Hamiltonian $H_0$, and continue to apply $GW$
approximation until self-consistency is achieved. The distinctive
feature of QS$GW$ compared with DFT+$U$ and DMFT is that it does not
require any {\it ad hoc} parameters.  Previous studies, ranging from atoms \cite{Bruneval}, 
semiconductors \cite{QSGW-2006, QSGW-2007, Shaltaf-PRL, Lambrecht} to the various 3$d$
transition metal oxides \cite{QSGW-2006, QSGW-2007, Kotani-JPCM,
  Han-nickelate, Jang-cuprate} and 4$f$ electron systems
\cite{Chantis-4f} have demonstrated its capacity to investigate 
 many different types of correlated materials.

\begin{figure} 
\includegraphics[width=0.45\textwidth, angle=0]{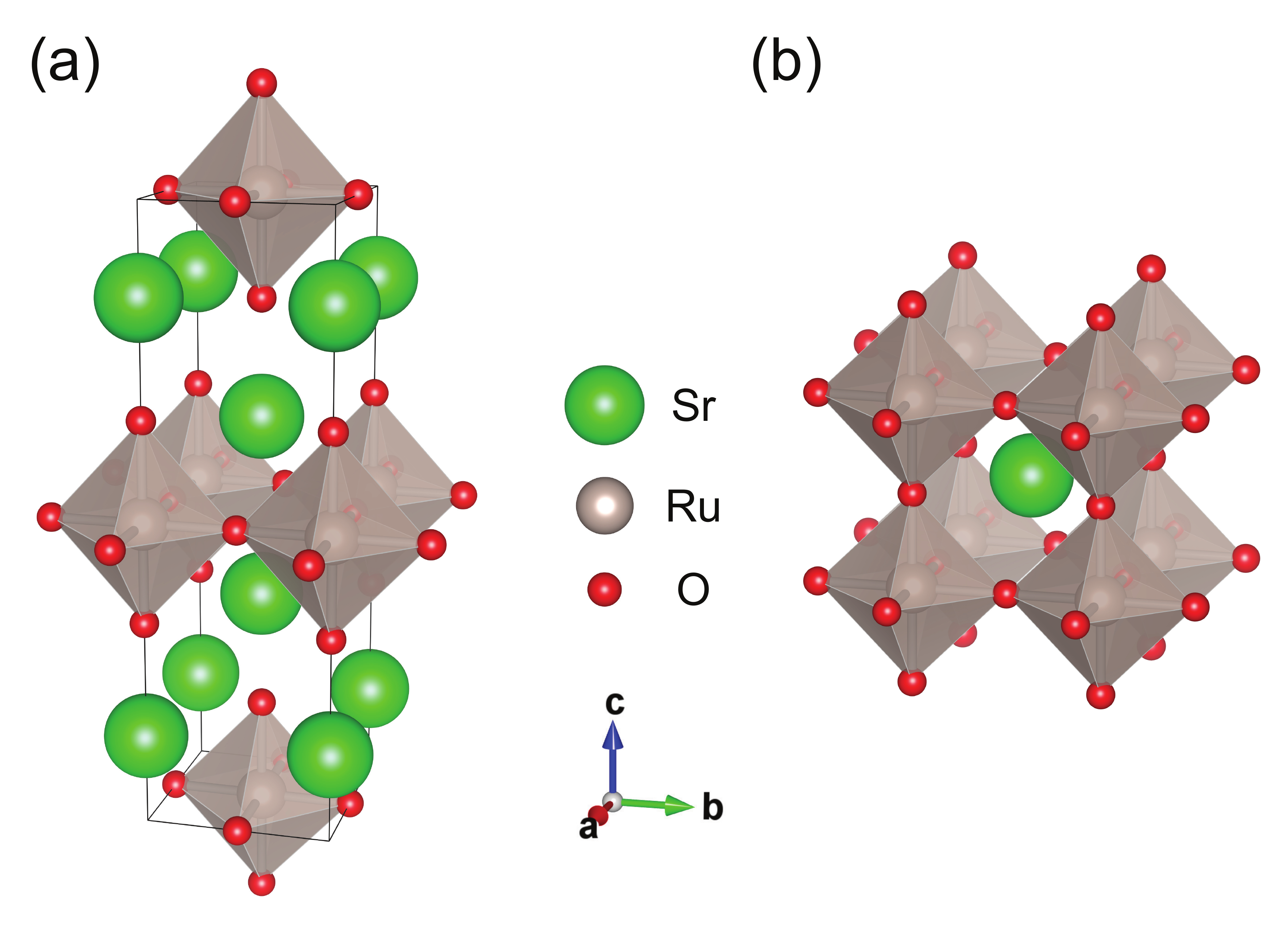}
\caption{The crystal structure of (a) SRO214 and (b) SRO113. The
  green, gray (octahedra), and red spheres represent Sr, Ru (RuO$_6$
  cage), and O atoms, respectively.}
\label{structure}
\end{figure}

\subsection{Computation details}
Our implementation of QS$GW$ in `ecalj' code \cite{ecalj-1}
is based on the `augmented plane wave (APW) + muffin-tin orbital
(MTO)' method, designated by `PMT' \cite{PMT}. The accuracy of this
scheme was proven to be satisfactory \cite{PMT}. A key feature of this
scheme for QS$GW$ is that the expansion of $V^{\rm xc}$ can be made
with MTOs, not APWs, which enables us to make the real space
representation of $V^{\rm xc}$ at any ${\bf k}$ point. In contrast to
the previous approach based on FP-LMTO (full potential linearized
muffin-tin orbital) \cite{QSGW-2007}, our scheme is free from the fine
tuning of MTO parameters.

We used the in-plane and out-of-plane lattice constant of 3.87 {\AA}
and 12.73 {\AA}, respectively, for the body-centered tetragonal
SRO214. For SRO113, the cubic perovskite structure is considered as
shown in Fig.~\ref{structure}(b). The pseudocubic lattice constant of
the orthorhombic structure is 3.93 {\AA} \cite{113-lattice}. We used
6$\times$6$\times$6 \cite{kpoints} and 8$\times$8$\times$8 {\bf k} points for
the self-energy calculation in the first Brillouin zone of SRO214 and
SRO113, respectively. For cubic SRO113 we also performed DFT+$U$
calculations using OpenMX code \cite{openmx} to make a comparison with
previous studies. Both LDA \cite{XC-LDA} and GGA \cite{XC-GGA}
functionals were used in combination with Dudarev form of DFT+$U$
\cite{Dudarev,Han+U}.  16$\times$16$\times$16 {\bf k} points were
adopted for DFT$+U$ calculations.

\begin{figure} 
\includegraphics[width=0.48\textwidth, angle=0]{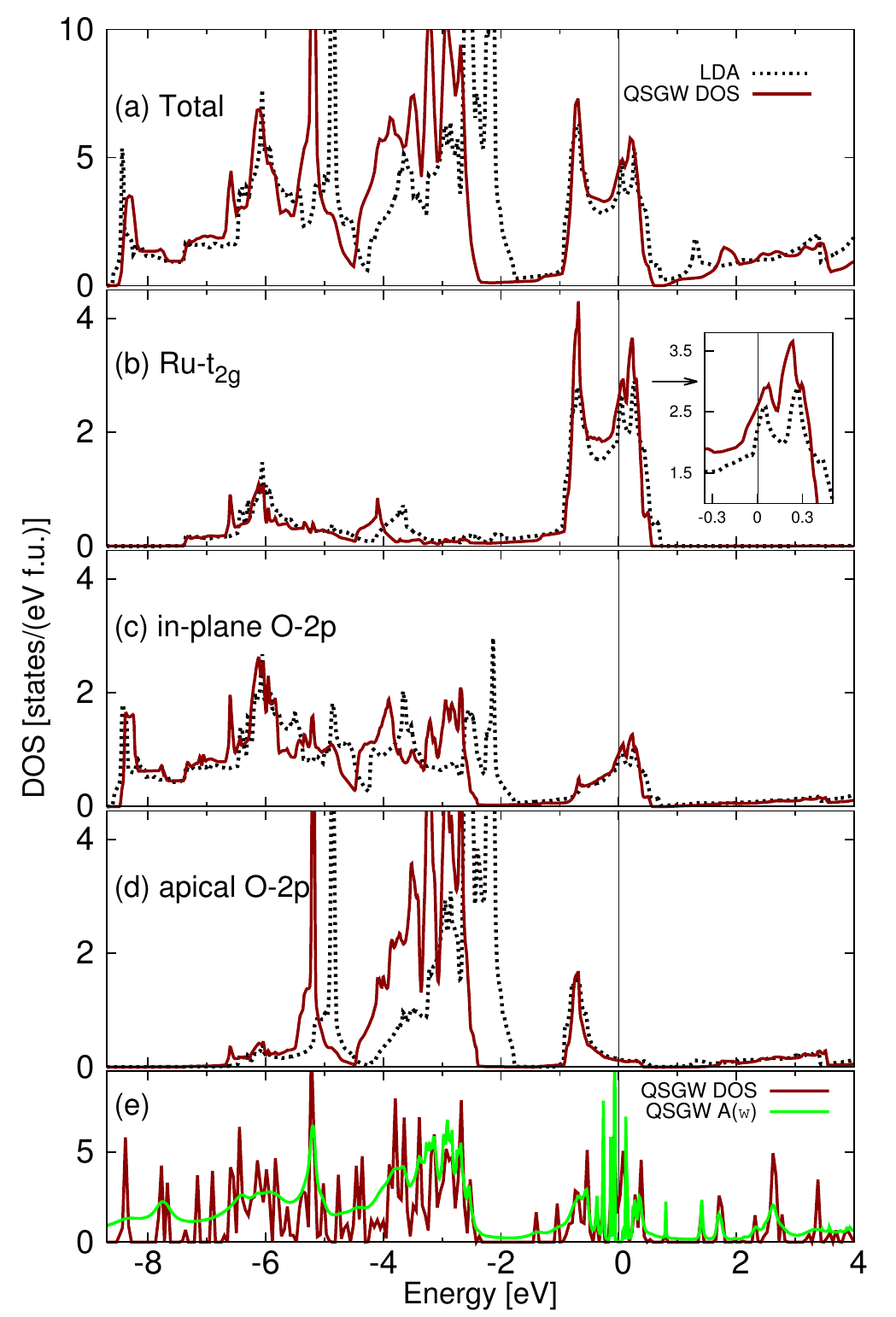}
\caption{The calculated (a) total DOS and (b--d) PDOS for SRO214 by
  LDA (black dotted lines) and QS$GW$ (red solid lines). The spectral function $A(\omega)$ 
calculated by QS$GW$ self-energy 
is plotted in (e) (green solid line) along with the non-interpolated total DOS (red solid line). The Fermi
  level is set to zero.}
\label{PDOS}
\end{figure}

\begin{figure} 
\includegraphics[width=0.45\textwidth, angle=0]{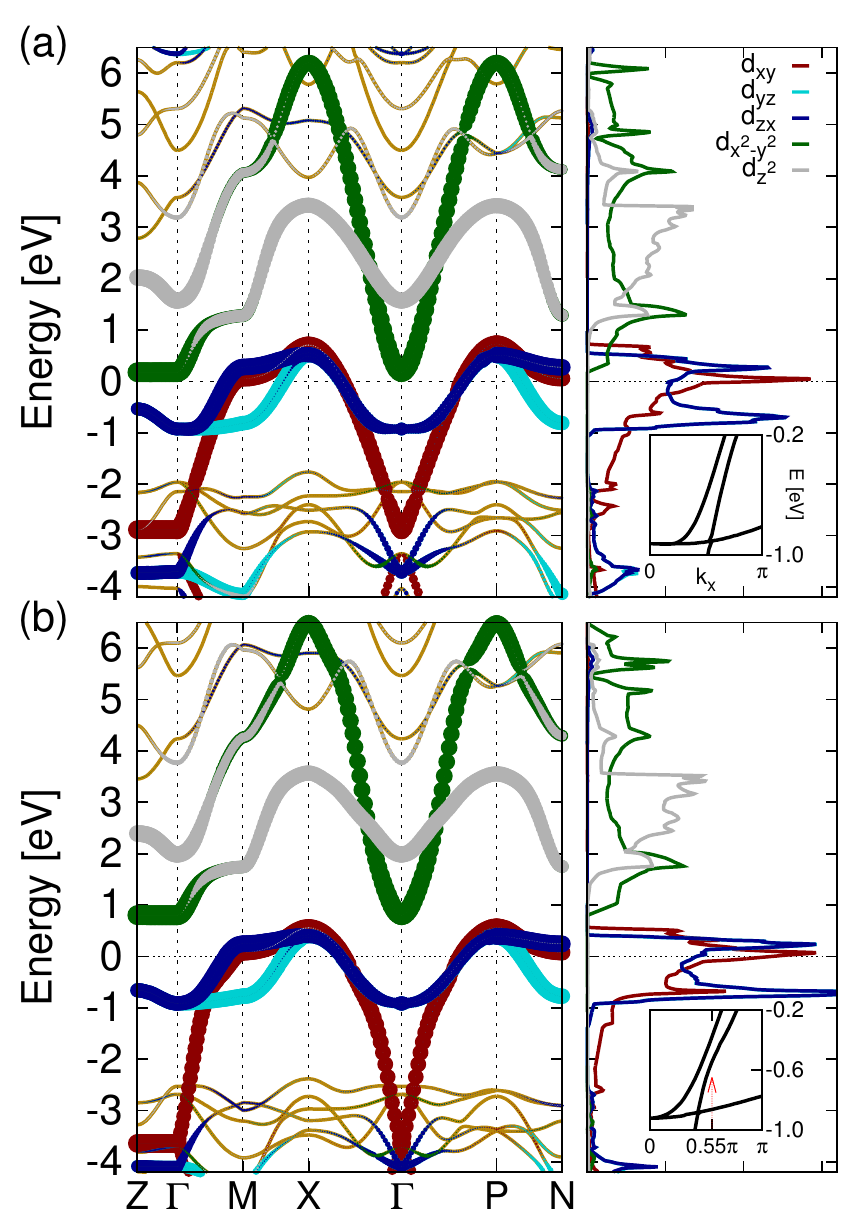}
\caption{The calculated band dispersion and the Ru-$d$ PDOS for SRO214
  by (a) LDA and (b) QS$GW$. The red, cyan, blue, green and gray lines
  refer to the Ru $d_{xy}$, $d_{yz}$, $d_{zx}$, $d_{x^2-y^2}$ and
  $d_{z^2}$ states, respectively. The yellow lines represent the O and
  Sr states (not shown in DOS). The thickness of the bands corresponds
  to the amount of the corresponding orbital character. The red and blue arrows represent the bandwidth of each band. The Fermi level is set to zero. The insets highlight the band dispersion at around $-0.6$ eV along $\Gamma$ to M ($\pi$,0,0) by (a) LDA and (b) QS$GW$. A sudden change of its slope at (0.55$\pi$,0,0) is observed only in QS$GW$ (red arrow). This feature is reflected in PDOS (see Fig.~\ref{t2g}(b)).}
\label{fat_band}
\end{figure}

\section{Result and discussion}
\subsection{Electronic structure of Sr$_2$RuO$_4$} \label{SRO214}
Our LDA result is in good agreement with the previous calculations
\cite{Oguchi-214, Singh-214, Hase-214, McMullan-214, Noce-214,
  Mackenzie-214}, see Fig.~\ref{PDOS} (black dotted lines) and
Fig.~\ref{fat_band}. The van Hove singularity (vHS) is located at
$\sim 60$ meV above the $E_F$ and the density of states (DOS) at $E_F$
is $N(E_F) \simeq 3.34$ states/(eV f.u.). The Ru-$t_{2g}$ state is dominant
near $E_F$ and is hybridized with O-2$p$. From Fig.~\ref{PDOS}(a), (c), and
(d), one can see that the total DOS in the range of $-4$ -- $-2$ eV is
largely attributed to the apical O-2$p$. The calculated band
dispersion is presented in Fig.~\ref{fat_band}(a). The bandwidth of
the 2D-like $d_{xy}$ state is about two times larger than that of
$d_{yz,zx}$ states. The nonbonding O-2$p$ state is located in $-4$ --
$-2$ eV as can also be seen in Fig.~\ref{PDOS}(d).

\begin{figure} 
\includegraphics[width=0.48\textwidth, angle=0]{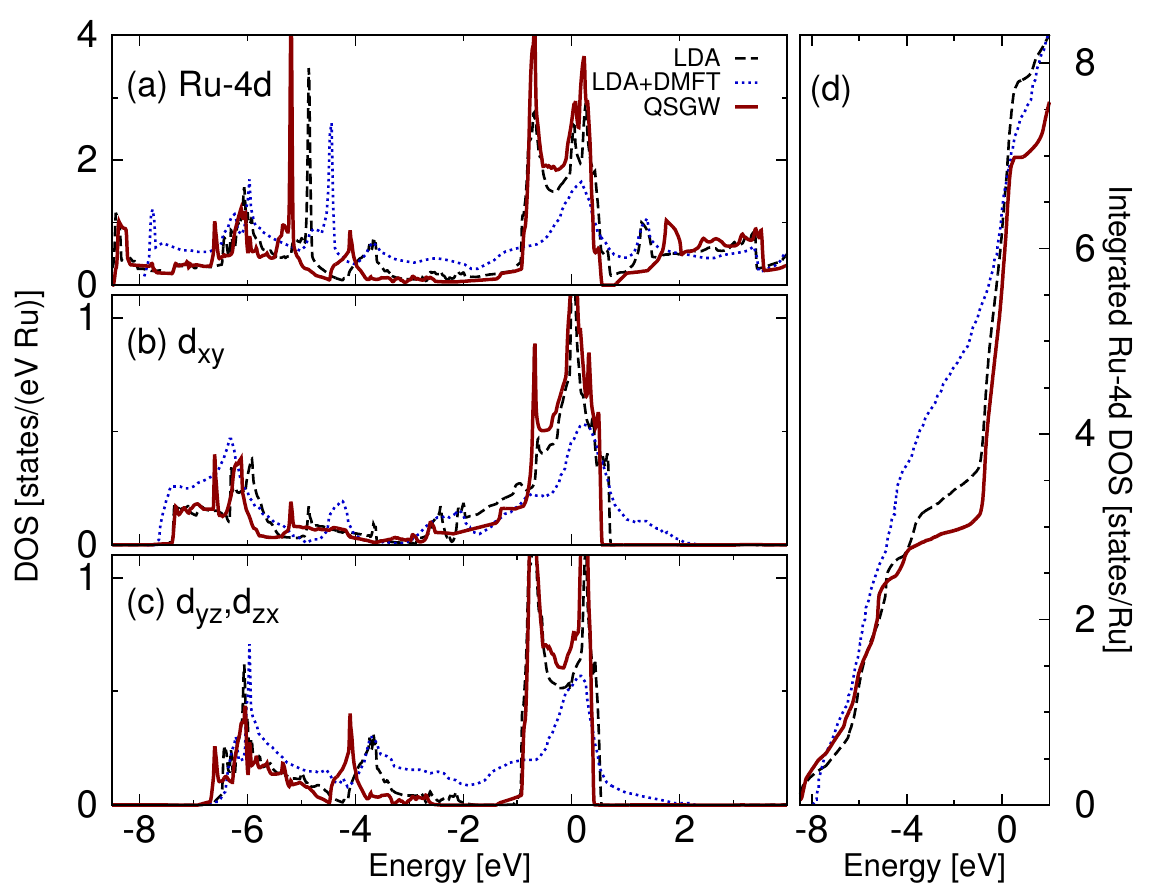}
\caption{(a--c) The PDOS for SRO214 calculated by LDA (black dashed
  lines), LDA+DMFT \cite{Pchelkina-DMFT} (blue dotted lines) and
  QS$GW$ (red solid lines). The $e_g$ states in the previous LDA and
  DMFT study were identical \cite{Pchelkina-DMFT} and are not shown
  here.  (d) The integrated $d$ DOS defined as
  $N(E)=\int_{-\infty}^{E} n(\varepsilon)d\varepsilon$, where
  $n(\varepsilon)$ is the DOS at energy $\varepsilon$. Each value was
  normalized to have an equal number of states at $E= \infty$. $E_F$
  is set to zero.}
\label{t2g}
\end{figure}

The QS$GW$ DOS is presented in Fig.~\ref{PDOS} (red solid
lines) \cite{spectral}. While the overall shape of the DOS is not much different from the
LDA results, some differences are observed. The two vHS located at $\sim +60$ and $\sim +260$ meV in LDA \cite{Oguchi-214, Singh-214, Hase-214, McMullan-214, Noce-214,
  Mackenzie-214} become closer in QS$GW$ shifting to $\sim +70$ and $\sim +220$ meV, respectively (inset in Fig.~\ref{PDOS}(b)).
Both the bonding (at $\sim -6$ eV) and the antibonding (around the
$E_F$) part of Ru-4$d$ states becomes flatter in their dispersion
compared to the LDA result. The $t_{2g}$
bandwidth is slightly reduced (see Fig.~\ref{PDOS} and \ref{fat_band}).
An interesting feature is found in the $d_{xy}$ band dispersion along $\Gamma$ to M. The different dispersion at $\sim -0.6$ eV (see Fig.~\ref{fat_band} insets) is related to the more pronounced peak in the QS$GW$ PDOS (Fig.~\ref{t2g}(b)).
The O-$2p$ states are shifted to the lower energy regions in QS$GW$ 
(Fig.~\ref{PDOS}(c)-(d)). The Sr-4$d$ state is further pushed up to
$\sim 4$ eV (not shown).

\begin{figure*}[t] 
\includegraphics[width=0.7\textwidth, angle=0]{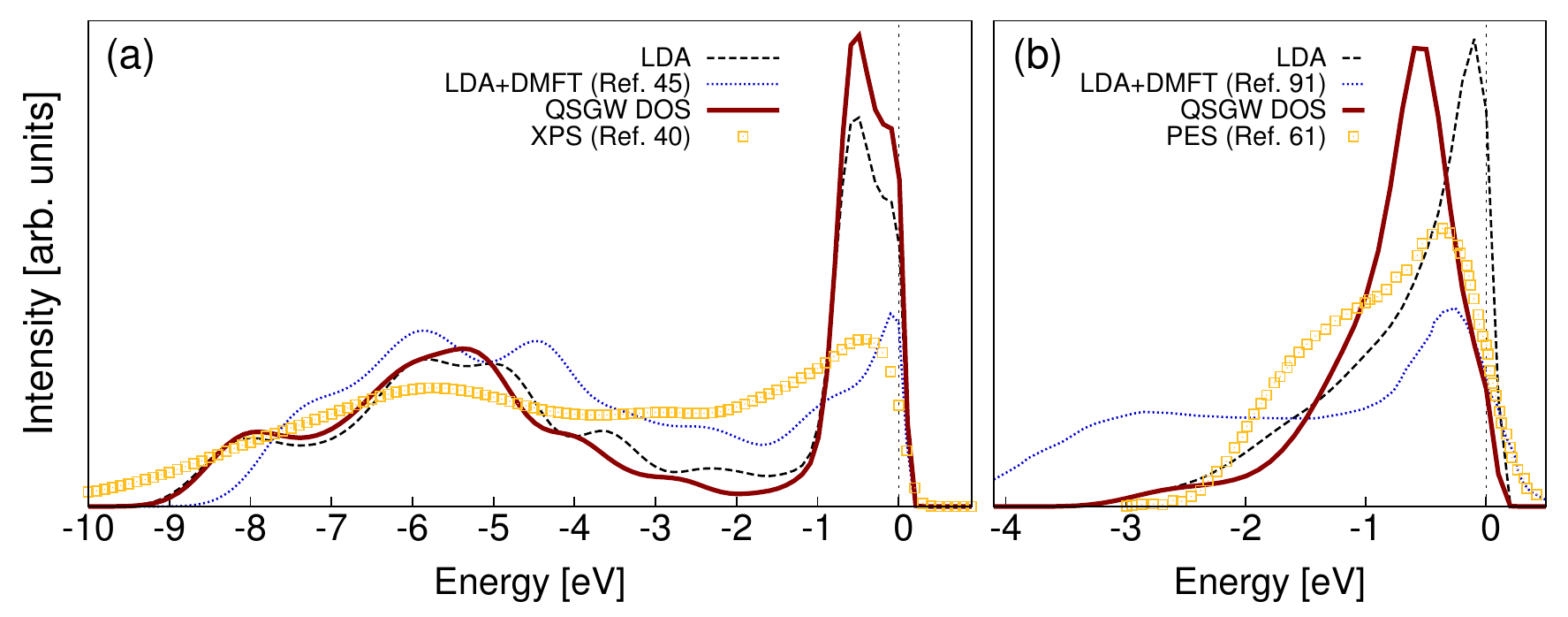}
\caption{The experimental and calculated spectra of (a) SRO214 and (b) Ru-4$d$ states of SRO113. In (a), XPS spectrum (yellow dots) by Yokoya {\it et al.} \cite{Yokoya-peak} are plotted along with Gaussian-broadened DOS by LDA (black dashed line), LDA+DMFT (blue dotted line) \cite{Pchelkina-DMFT}, and QS$GW$ (red solid line). In (b), PES spectrum (yellow dots) by Takizawa {\it et al.} \cite{Takizawa} obtained from orthorhombic SRO113 at room temperature are plotted with Gaussian-broadened DOS of cubic SRO113 (LDA and QS$GW$) and orthorhombic SRO113 (LDA+DMFT \cite{Jakobi-113DMFT}). Only Ru-4$d$ states are plotted in (b). The intensities were normalized in the area under curves.} 
\label{spectra}
\end{figure*}

The further details of Ru-$t_{2g}$ state is of particular interest as
the LHB feature at around $-$3 eV has been a subject of debate
\cite{Liebsch-DMFT, Pchelkina-DMFT, Singh-comment, Pchelkina-reply,
  Yokoya-peak, Tran-peak}. Fig.~\ref{t2g} compares the calculated PDOS
by LDA, LDA+DMFT, and QS$GW$. Note that the states in the range of
$-$4 -- $-$1 eV is noticeably larger in LDA+DMFT than LDA and QS$GW$
especially for $d_{yz,zx}$. It is related to the spectral weight
transfer from the near-$E_F$ regions to the LHB. This is one of the
main features of DMFT calculations and was the main point of the
previous debate between LDA, DMFT \cite{Pchelkina-DMFT,
  Singh-comment, Pchelkina-reply} and the XPS experiment \cite{Yokoya-peak}, while The QS$GW$ result in this region is
similar with the LDA rather than the DMFT. Namely, QS$GW$ procedure
does not capture the dynamic `Mott-Hubbard physics' well as in DMFT
while it still takes some correlation effect into account as reflected
in the bandwidth reduction and the mass enhancement. This can also be seen in Fig.~\ref{spectra}(a). One can notice the shoulder-like LHB feature in XPS spectrum at $\sim -$ 3 eV being consistent with LDA+DMFT, while both LDA and QS$GW$ give the less states in this energy region. The theoretical spectra in Fig.~\ref{spectra}(a) were broadened as in Ref.~\cite{Pchelkina-DMFT}. Not surprisingly,
this LHB feature of DMFT is more pronounced in $d_{yz,zx}$ states than
in $d_{xy}$ because of the narrower bandwidth. This point is
highlighted in the integrated DOS (Fig.~\ref{t2g}(d)). While the total
number of Ru states (the integrated value up to $E_F$) is basically the
same in all three calculations, the DMFT value gets increased in a
wider energy range.  On the other hand, LDA and QS$GW$ values are much
more rapidly increased in the narrower energy range of $-0.5 \leq
\varepsilon \leq 0$ eV. Since the QS$GW$ bandwidth is narrower than
LDA, this increase is more pronounced in QS$GW$. Our result
demonstrates the characteristic feature of QS$GW$ to take into account
of electron correlations distinctive from DMFT  
especially regarding the incoherent states at $-3$ eV.

The Ru-$e_g$ bands are also affected by QS$GW$ self-energy (see
Fig.~\ref{fat_band}). The slight up-shift of the $e_g$ anti-bonding
bands is consistent with the previous study of SrVO$_3$
\cite{Tomczak-bandwidening}. As a result, the $d_{x^2-y^2}$ band does
not touch the $E_F$ even at $\Gamma$ and $Z$ point (see
Fig.~\ref{fat_band}). The O-2$p$ nonbonding state is located at $-3.5$
-- $-2.5$ eV in QS$GW$ (Fig.~\ref{fat_band}), which is in better
agreement with the recent ARPES result \cite{Iwasawa-ARPES} than LDA.

\begin{figure} 
\includegraphics[width=0.27\textwidth, angle=0]{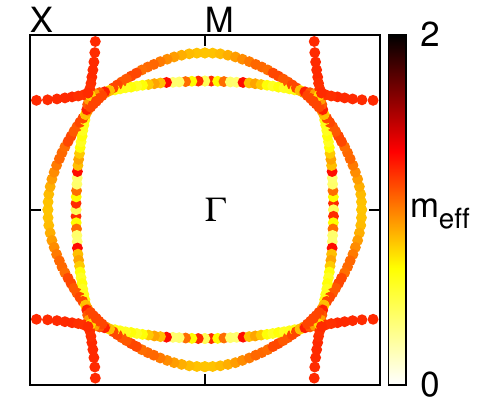}
\caption{The Fermi surface of SRO214 at $k_z = 0$ plane. The color
  represents the calculated effective mass defined as $m_{\rm QSGW}/m_{\rm LDA}$.}
\label{FS_mass}
\end{figure}

The overall shape of the calculated Fermi surface by QS$GW$ (see Fig.~\ref{FS_mass}) is consistent with the LDA \cite{Oguchi-214, Singh-214, Hase-214, McMullan-214, Noce-214, Mackenzie-214}, dHvA \cite{Mackenzie-mass, Bergemann-dHvA} and ARPES \cite{Damascelli-FS, Matzdorf-Science, Shen-ARPES}. The inclusion of spin-orbit coupling as a perturbative correction within QS$GW$ induces relatively small modifications at the band crossing points (not shown) \cite{Haverkort-SOC}. The calculated effective mass at each {\bf k} direction is
shown in Fig.~\ref{FS_mass} with color plot where  $m^{*}/m_{\rm LDA}$ is
estimated simply by taking the derivative of band dispersions. Along
the $\Gamma$ to X ($\pi$,$\pi$,0) line, the mass enhancement by QS$GW$
is about 15\% ($m^{*}_{xy}/m_{\rm LDA} \simeq 1.15$) for $d_{xy}$ and
22\% ($m^{*}_{yz,zx}/m_{\rm LDA} \simeq 1.22$) for $d_{yz,zx}$,
respectively. For the $\Gamma$ to M ($\pi$,0,0) line,
 $m^{*}_{xy}/m_{\rm LDA} \simeq 0.83$ and
$m^{*}_{yz,zx}/m_{\rm LDA} \simeq 0.78$.

It is instructive to compare the QS$GW$ result with the DMFT calculations
and experiments 
while the DMFT effective mass strongly depends on the $U$ and $J$
values. The value reported by Mravlje {\it et al.}  with $U=1.7$ and
$J=0.0$ -- 0.1 eV is $m^{*}/m_{\rm LDA} \simeq 1.7$ for both $d_{xy}$
and $d_{yz,zx}$ \cite{Mravlje-DMFT}. This value is significantly smaller than the other
DMFT calculations, {\it e.g.}, by Pchelkina {\it et al.},
$m^{*}/m_{\rm LDA}$ $\simeq 2.6$ and 2.3 for $d_{xy}$ and $d_{yz,zx}$,
respectively (with $U=3.1$ and $J=0.7$ eV obtained from constrained
LDA) \cite{Pchelkina-DMFT}, and by the same group but with $U=2.3$ and
$J=0.2$ eV, $m^{*}/m_{\rm LDA}$ is $\simeq 2.3$ and 2.0 for $d_{xy}$
and $d_{yz,zx}$, respectively \cite{Mravlje-DMFT}. These DMFT values
were calculated from $m_{\rm DMFT}/m_{\rm LDA} \equiv Z_{\rm
  DMFT}^{-1} = [1-\partial_{\omega}{\rm Re}\Sigma_{\rm
    DMFT}(\omega)]_{\omega = 0}$. Our estimation from QS$GW$
self-energy yields $Z_{\rm QSGW}^{-1} \simeq 1.82$ for $d_{xy}$ and
1.71 for $d_{yz,zx}$ \cite{Z-factor} which is not much different from
the DMFT values of 1.7 calculated by Mravlje {\it et al.} with
$U=1.7$ and $J=0.0$ -- 0.1 eV \cite{Mravlje-DMFT}. 
Note that $U=2.3$ and $J=0.4$ eV were suggested to be reasonable \cite{Mravlje-DMFT} 
in comparison to dHvA measurements \cite{Mackenzie-mass, Bergemann-mass}, and that 
the experimental values are larger than the QS$GW$. The early ARPES
reports $m^{*}/m_{\rm LDA} \simeq 2.5$ \cite{Puchkov-mass} and the
more recent measurement by Iwasawa {\it et al.} is $m^{*}_{xy}/m_{\rm
  LDA} \simeq 3.7$ and $m^{*}_{yz,zx}/m_{\rm LDA} \simeq 2.0$
\cite{Iwasawa-ARPES}. There are two reports from the dHvA experiment;
$m^{*}_{xy}/m_{\rm LDA} \simeq 4.1$ and $m^{*}_{yz,zx}/m_{\rm LDA}
\simeq 3.3$ by Mackenzie {\it et al.}  \cite{Mackenzie-mass}, and
$m^{*}_{xy}/m_{\rm LDA} \simeq 5.5$ and $m^{*}_{yz,zx}/m_{\rm LDA}
\simeq 3.4$ by Bergemann {\it et al.}  \cite{Bergemann-mass}.

\subsection{Electronic structure of cubic SrRuO$_3$} \label{SRO113}

\begin{figure} 
\includegraphics[width=0.48\textwidth, angle=0]{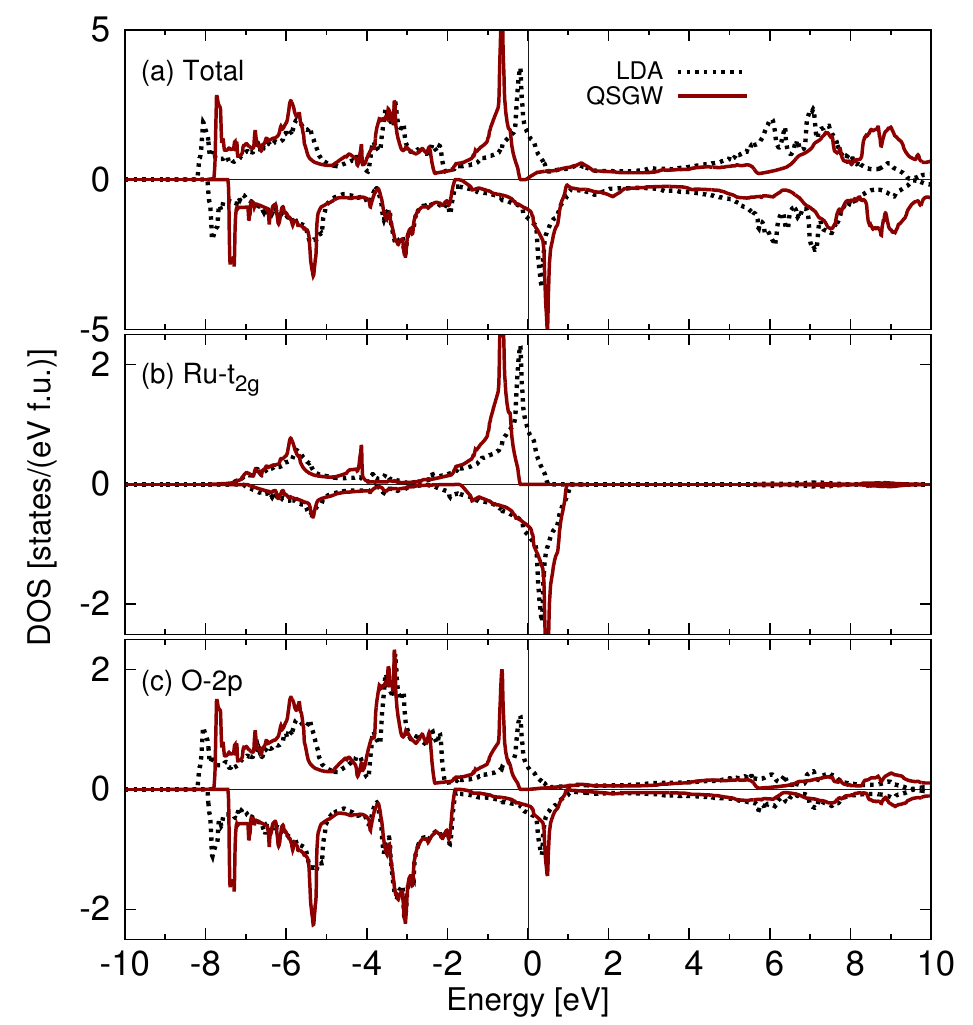}
\caption{The calculated (a) total DOS and (b--c) PDOS of cubic SRO113
  by LDA (black dotted lines) and QS$GW$ (red solid lines). The Fermi
  level is set to zero. The positive and negative DOS represent the
  up- and down-spin parts, respectively.}
\label{113_PDOS}
\end{figure}

\begin{figure} 
\includegraphics[width=0.48\textwidth, angle=0]{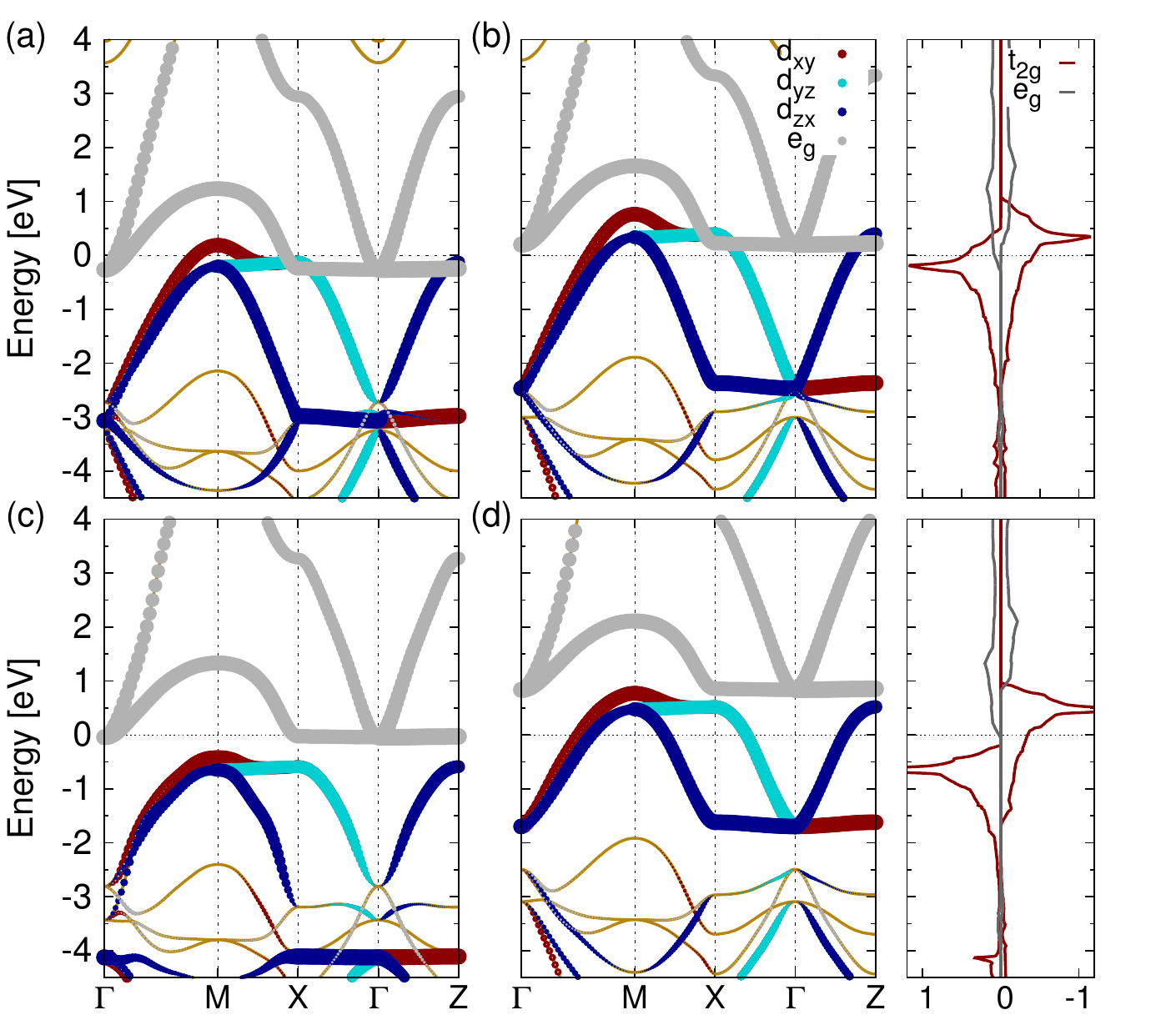}
\caption{The calculated band dispersion and the Ru $d$-orbital PDOS for
  cubic SRO113 by (a) LDA majority spin band, (b) LDA minority spin
  band, (c) QS$GW$ majority spin band, and (d) QS$GW$ minority spin
  band. The red, cyan, blue, and gray lines refer the Ru $d_{xy}$,
  $d_{yz}$, $d_{zx}$, and $e_g$ states, respectively. The yellow lines
  represent the O and Sr states (not shown in DOS). The thickness of
  the bands corresponds to the amount of the corresponding orbital
  character. The Fermi level is set to zero. }
\label{113_fat_band}
\end{figure}

The LDA result for cubic SRO113 is in good agreement with the previous
studies \cite{Allen-113, Santi-113, Mazin-113, Rondinelli-113} (see
Fig.~\ref{113_PDOS} and Fig.~\ref{113_fat_band}). The calculated
magnetic moment of $\mu = 1.28 \mu_{\rm B}$/f.u. also reasonably well
agrees with the literature values of $\mu = 1.09 \mu_{\rm
  B}$/f.u. (calculated by VASP) and $\mu = 1.26 \mu_{\rm
  B}$/f.u. (calculated by SIESTA) \cite{Rondinelli-113}.  The Sr-4$d$
state is located above $\sim 4$ eV (not shown) and the nonbonding
state of O-2$p$ is in $-4$ -- $-2$ eV. The antibonding Ru-$t_{2g}$
character dominates the near-$E_F$ region and the bonding complex is
located at $-8$ -- $-4$ eV. A clear splitting between the up and down
spin DOS is noticed and is responsible for the ferromagnetism in this
material. The effect of orthorhombic distortion (not taken into account in our calculation) has been investigated previously.
For example, Rondinelli {\it et al.} \cite{Rondinelli-113} showed that
 the result from orthorhombic structure is similar to the cubic case
other than the slightly reduced exchange splitting at $\Gamma$ point and bandwidth reduction by $\sim 0.35$ eV for $t_{2g}$, 1.5 eV for $e_g$, and 0.6 eV for O 2$p$.

\begin{figure} 
\includegraphics[width=0.48\textwidth, angle=0]{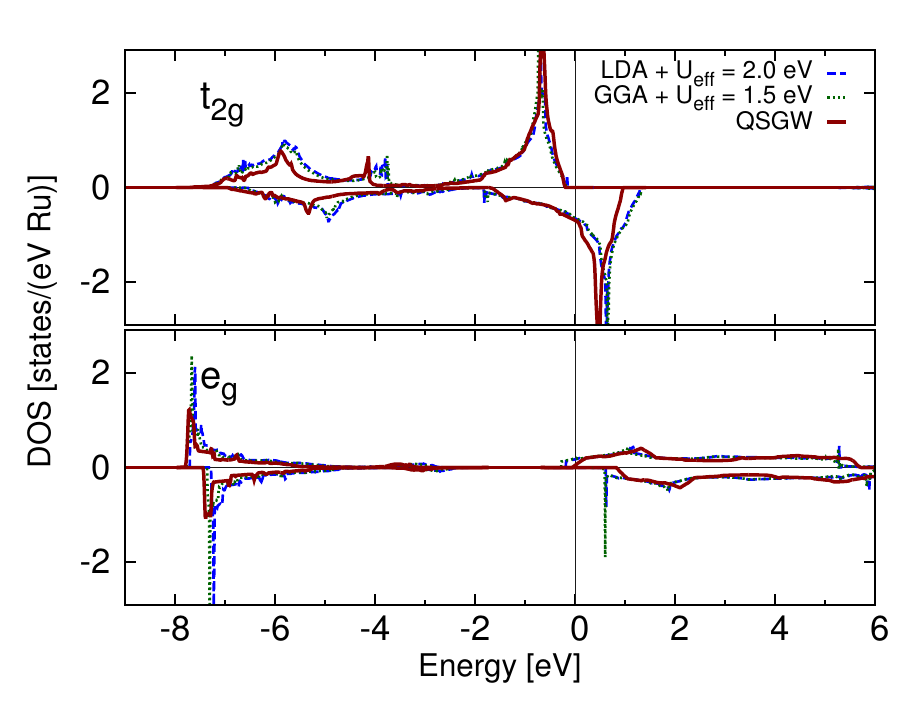}
\caption{The calculated Ru $t_{2g}$ and $e_g$ PDOS of cubic SRO113 by
  LDA+$U$ ($U_{\rm eff}=2.5$ eV) (blue dashed lines), GGA+$U$
  ($U_{\rm eff}=2.0$ eV) (green dotted lines) and QS$GW$ (red solid
  lines). The Fermi level is set to zero. The positive and negative
  DOS represent the up- and down-spin parts, respectively.}
\label{113_U}
\end{figure}

Several distinctive features are found in the QS$GW$ results. First,
the notable bandwidth reduction is observed (see Fig.~\ref{113_PDOS} and
Fig.~\ref{113_fat_band}). The majority spin bandwidth is reduced by
$\sim 0.7$ eV (from $\sim 3.2$ eV (LDA) to $\sim 2.5$ eV (QS$GW$)) and
the minority spin bandwidth is by $\sim 0.6$ eV (from $\sim 3.2$ eV
(LDA) to $\sim 2.6$ eV (QS$GW$)).  The
exchange splitting is enhanced to be $\sim 1.2$ eV which is
significantly larger than the LDA value of $\sim 0.5$ eV. We found the
naive comparison of the effective mass for LDA and QS$GW$ based on the
{\bf k}-derivative can be misleading in SRO113 due to the different
Fermi wave vectors caused by the enhanced exchange splitting in QS$GW$.
The calculated $Z_{\rm QSGW}^{-1}$ for the up and down spin is 1.33
and 1.68, respectively. 
The effective mass of $m_{\rm QSGW}/m_{\rm LDA} = 1.26$ estimated from the bandwidth ratio is reasonably well compared with some of
DMFT results while the DMFT values show significant
deviations ranging from 1.1 to 4.5 \cite{Jakobi-113DMFT,Granas-2014,Kim-Min,Dang-Hund}.  
The specific heat measurements report the mass enhancement of $m^{\rm*}/m_{\rm LDA} = 3.7$ \cite{Allen-113} and 4.5 \cite{Okamoto-spec}. 
The discrepancy reflects not only the limitation of RPA-QS$GW$ correlations but also the effect of orthorhombic distortion \cite{Takizawa, Pavarini,Georges-Hund}.

Together with the bandwidth reduction and the enhanced exchange
splitting, the QS$GW$ electronic structure becomes half-metallic with
a gap in the majority spin state.  Note that this QS$GW$ band
structure is quite similar with DFT$+U$ result \cite{Jeng-half,
  Mahadevan-SRO} as shown in Fig.~\ref{113_U}. The calculated magnetic
moment of QS$GW$ is $\mu = 2.0 \mu_{\rm B}/$f.u. which is
significantly larger than the LDA value and comparable with the
DFT$+U$ results \cite{Jeng-half, Mahadevan-SRO, Rondinelli-113}. 
While we only considered the cubic structure, the effect of 
orthorhombic distortion is basically to enhance the on-site correlations as reported
by the previous study \cite{Takizawa, Pavarini}. 
The experimental verification of the half-metallicity may not be easy
because of the large magnetic fields required to overcome the magnetic
anisotropy \cite{Verissimo-SRO}. This possibility in SRO113 has been
actively discussed based on DFT$+U$ calculations \cite{Jeng-half,
  Mahadevan-SRO} and hybrid functional \cite{Verissimo-SRO}. Therefore
our QS$GW$ result adds a new promising aspect toward this direction.

The detailed comparison with DFT$+U$ result is given in
Fig.~\ref{113_U}. The optimized values of $U_{\rm eff}$ are favorably
compared with a recent constrained RPA result of $U_{\rm eff} = 2.1$ eV although
it is calculated from the orthorhombic structure \cite{Mahadevan-SRO}.
For comparison one can recall that SIC-LDA result is reproduced by
$U_{\rm eff} = 1$ eV in the case of orthorhombic structure of SRO113
\cite{Rondinelli-113}. As for the DMFT calculations, several different
choices of $U$ and $J$ have been made as in the case of SRO214. In
terms of $U_{\rm eff}$=$U-J$, it ranges from 1.75 to 2.4 eV
\cite{Jakobi-113DMFT,Granas-2014,Dang-Hund}. 

Several photoemission spectroscopy (PES) experiments report the LHB-like feature at $\sim -$2 eV \cite{Fujioka-spec, Okamoto-spec, Takizawa}. The PES spectrum of orthorhombic phase \cite{Takizawa} are plotted along with the calculated DOS by LDA, LDA+DMFT \cite{Jakobi-113DMFT}, and QS$GW$ in Fig.~\ref{spectra}(b). The LHB-like peak near $-$ 1.5 eV is observed in the PES spectra, while not in LDA and QS$GW$. It has been argued that this LHB-like feature is not related to the magnetic fluctuations or the orthorhombic structural distortion \cite{Takizawa}. LDA+DMFT calculation of orthorhombic SRO113 with $U=3.5$ and $J=1.75$ eV predicts the LHB at $-3$ -- $-1$ eV \cite{Jakobi-113DMFT} (Fig.~\ref{spectra}(b)). As in SRO214, LHB is the characteristic feature of DMFT distinctive from QS$GW$ (also from DFT$+U$) while some DMFT results do not seem to support this feature \cite{Dang-Hund}. The absence of this state in QS$GW$ and
DFT$+U$ (see $t_{2g}$ states of Fig.~\ref{113_U}) indicates that it is related to the dynamic aspects of Ru-$4d$
correlation hybridized with O-$2p$ as in the case of Zhang-Rice band
in cuprates. Also, regarding the half-metallic band structure, DMFT
calculations do not seem to give a consistent prediction
\cite{Jakobi-113DMFT,Kim-Min}. These issues that is likely related to
the intrinsic parameter dependency require further investigations.

\section{Summary} 
Using QS$GW$ calculations, we re-investigated the electronic structure
of SRO214 and SRO113. Without any {\it ad hoc} parameter or the
ambiguity related to the double-counting and downfolding issues, some
of the important features of electron correlations were reasonably
well captured such as the bandwidth renormalization and the exchange
splitting. In the case of SRO113, QS$GW$ result is in good agreement
with DFT+$U$ in a reasonable range of $U$ and $J$ parameter. While the
QS$GW$ shows the limitation in describing the detailed features such
as the Ru-$4d$ spectral weight transfer to LHB, it can be improved in
combination with other techniques as reported in recent studies
\cite{QSGW+DMFT, Tomczak-bandwidening, Choi-2015}. Our result sheds
new light on the possibility and the limitation of the
first-principles electronic structure calculations of the moderately
correlated transition-metal oxides systems.

\section{Acknowledgements}
S.R., S.W.J., and M.J.H. were supported by Basic Science Research
Program through the National Research Foundation of Korea (NRF)
(2014R1A1A2057202; Ministry of Education) and by Samsung Advanced
Institute of Technology (SAIT). The computing resource is supported by
National Institute of Supercomputing and Networking / Korea Institute
of Science and Technology Information with supercomputing resources
including technical support (KSC-2013-C2-024) and by Computing System
for Research in Kyushu University. T.K. was supported by the Advanced
Low Carbon Technology Research and Development Program (ALCA), the
“High-efficiency Energy Conversion by Spinodal Nano-decomposition”
program of the Japan Science and Technology Agency (JST), and the JSPS
Core-to-Core Program Advanced Research Networks (''Computational
Nano-materials Design on Green Energy'').


\begin{thebibliography}{99}
\bibitem{spectral} There is no significant difference between DOS and $A(\omega)$ obtained by QS$GW$ (see Fig.~\ref{PDOS}(e)). Therefore
in the below, our discussion will mainly be based on DOS.
\end{thebibliography}
\end{document}